\documentclass[twocolumn,showpacs,preprintnumbers,amsmath,amssymb]{revtex4}

\usepackage{graphicx}
\usepackage{dcolumn}
\usepackage{bm}

\newcommand{\beq}{\begin{eqnarray}}
\newcommand{\eeq}{\end{eqnarray}}
\newcommand{\beqq}{\begin{eqnarray*}}
\newcommand{\eeqq}{\end{eqnarray*}}

\begin{document}
\title{A quantum decay model with exact explicit analytical solution
}
\author{Avi Marchewka} \email{Avi_marchewka@yahoo.co.uk}
\author{Er'el Granot} \email{erel@yosh.ac.il}
\affiliation{Department of Electrical and Electronics Engineering,
College of Judea and Samaria, Ariel, Israel}




\begin{abstract}
A simple decay model is introduced. The model comprises of a point
potential well, which experiences an abrupt change. Due to the
temporal variation the initial quantum state can either escape
from the well or stay localized as a new bound state. The model
allows for an exact analytical solution while having the necessary
features of a decay process. The results show that the decay is
never exponential, as classical dynamics predicts. Moreover, at
short times the decay has a \textit{fractional} power law, which
differs from perturbation quantum methods predictions.
\end{abstract}


\maketitle

Despite the fact that almost all the decay processes in nature
observed of having an exponential decay law, it is well known that
quantum mechanics predicts deviation from this law
\cite{Fondaetal_1,Greenland_2,Chiuetal_3}.

In fact, it has been proven by Khalfin \cite{Khalfin_4} that the
long time behavior of the non-decay probability \textit{cannot} be
exponential, and in practical cases it has a negative power decay
law. Moreover, at short times most quantum systems obey $t^2$
dependence. This kind of dependence is usually attributed to a
reversible process, which contradicts an irreversible decay to the
continuum.

Recently, it became technologically feasible to measure this
deviation from exponential law, and indeed it was demonstrated
experimentally \cite{Menonetal_5,Wilkinsonetal_6,Grenland_7}.

The scenario of an irreversible decay of a confined bound state to
the continuum is ubiquitous in the physical world, Beta decay is
such a realization. Hence, resolving this controversy is required
to the understanding of these basic processes. Theoretically, this
problem was confronted by applying an abrupt perturbation on the
initial confined state. However, this perturbation approach merely
emphasizes the controversy except for the intermediate times where
an approximately exponential law seems to appear
\cite{CohenTannoudjietal_8}. At short times the reminiscent of a
reversible $t^2$ law is still dominant.

The recent technological developments, which allow trapping cold
atoms in very small traps \cite{Szriftgiseretal_9,Fortetal_10},
also allow to release them almost instantaneously (since the
trapping is done by laser beams). As a consequence, the temporal
dynamics and decay of quantum particles at the presence of an
abruptly changing potential can also be investigated in the
laboratory. In this paper we investigate a simple quantum
mechanical model, which can emulate realistic decay scenarios. The
initial state is a bound eigenstate of a localized potential well,
and the model allows investigating the state dynamics due to
abrupt change in this potential well. To simplify the model we use
a delta-function potential well. It is well known that this kind
of a potential can emulate \textit{any} barrier/well whose
de-Broglie wavelength is considerably longer than the potential
physical dimensions \cite{Granot_11}. In other words, for most
practical purposes it can replace any point potential. On the
other hand, a point potential brings in singularity to the system
and as a result a fractional power law emerges. However, as was
mentioned elsewhere \cite{GranotMarchewka_12} some reminders to
this behavior can be traced even in the analytic potential case.

The main strength of this model is that it has an exact analytical
solution and no approximations are taken. We show that the
dynamics of this model does not look exponential at any time.
Moreover, at short times (as well as at very long times) the
dynamics is governed by a \textit{fractional} power law instead of
an integral power law. To the best of our knowledge, this is the
only quantum decay model, where the final state can be either
localized or extended, and has an \textit{exact analytical}
solution.

We use a delta function to simulate the attractive potential well
\begin{equation}
\label{eq1} V\left( {x,t} \right) = \left\{
{{\begin{array}{*{20}c}
 { - 2\alpha \delta \left( x \right)} \hfill & {\mbox{for}} \hfill & {t \le
0} \hfill \\
 { - 2\lambda \delta \left( x \right)} \hfill & {\mbox{for}} \hfill & {t >
0} \hfill \\
\end{array} }} \right.
\end{equation}

That is, the potential can be written as an initial stationary
potential $V_0 \left( x \right) = - 2\alpha \delta \left( x
\right)$ and a perturbation part $\Delta V\left( {x,t} \right) =
2\left( {\alpha - \lambda } \right)u\left( t \right)\delta \left(
x \right)$ [$u\left( x \right)$ is the heaviside function $u\left(
x \right) = \left\{ {\mbox{0 for }x < 0\mbox{; 1 for }x > 0}
\right\}$]. The Schr\"{o}dinger equation reads

\begin{equation}
\label{eq2} i\frac{\partial \psi }{\partial t} = - \frac{\partial
^2\psi }{\partial x^2} + V_0 \left( x \right)\psi + \Delta V\left(
{x,t} \right)\psi
\end{equation}
\noindent hereinafter we adopt the units $\hbar = 1$ (Planck
constant) and $2m = 1$(particle mass). It should be stressed that
any shallow potential well ($\Delta x >> \left( {U_0 } \right)^{ -
1 / 2}$with width $\Delta x$ and depth $ - U_0 )$ can be replaced,
for most practical purposes, by a delta function potential
$V_{initial} \left( x \right) = - 2\alpha \delta \left( x \right)$
with the prefactor $2\alpha = U_0 \Delta x$, since the two
scatterers have a very similar scattering and a single bound state
( for a positive potential see, for example \cite{Granot_11}).

By applying the same analysis and logic of
\cite{GranotMarchewka_12} it can be shown that when the well has
finite width (instead of a delta function), say $\Delta x$, then
all the derivations that follows are valid provided $t>>\Delta
x^2$. Therefore, this behavior can be traced for \emph{every}
shallow potential well, i.e., where its eigen-boundstate energy
$E_0$ obeys $|E_0|<<\Delta x^{-2}$.

Next, for simplicity, we renormalize space and time to the
dimensionless variables $x_{new} = \alpha x, \quad t_{new} =
t\alpha ^2$and choose the normalized parameter $\mu \equiv \lambda
/ \alpha $. With these dimensionless parameters, the potential can
be written \[ V\left( {x,t} \right) = \left\{
{{\begin{array}{*{20}c}
 { - 2\delta \left( x \right)} \hfill & {\mbox{for}} \hfill & {t \le 0}
\hfill \\
 { - 2\mu \delta \left( x \right)} \hfill & {\mbox{for}} \hfill & {t > 0}
\hfill \\
\end{array} }} \right.
\]

Therefore, if the initial state $\psi _i \left( {x,t} \right)$ is
the bound eigenstate $\psi _{Bi} \left( {x,t} \right)$ of the
unperturbed well, then $ \label{eq3} \psi _i \left( {x,t} \right)
= \psi _{Bi} \left( {x,t} \right) \equiv \exp \left( { - \left| x
\right| + it} \right) $. After the abrupt potential change, the
only localized state is, of course
\begin{equation}
\label{eq4} \psi _{Bf} \left( {x,t} \right) = \mu ^{ - 1 / 2}\exp
\left( { - \mu \left| x \right| + i\mu ^2t} \right).
\end{equation}

This model has an exact solution without the need for any
simplifying approximations. The wavefunction at \textit{any}
instant can be calculated by the integral expression $ \label{eq5}
\psi \left( {x,t} \right) = \int\limits_{ - \infty }^\infty
{K\left( {x,x';t} \right)\psi _i \left( {x',t = 0} \right)} dx' $
where the Kernel of the integral (the Green function) is
[13]
 \beqq
&&K\left({x,x';t}\right)= K_{free} \left( {x,x';t} \right)+\\
&&\frac{\mu }{2}\exp \left[{-\mu \left( {\left| x \right| + \left|
{x'} \right|- i\mu t}\right)}\right]{\mbox{erfc}}\left(
{\frac{\left| x \right| + \left| {x'} \right| - i2\mu t}{2\sqrt
{it}}}\right)
\eeqq

 (where $\mbox {erfc}$ is the complementary error function \cite{AbramowitzStegun_14}) and the
free space Kernel is

$ K_{free} \left( {x,x';t} \right) \equiv \frac{1}{2\sqrt {i\pi t}
}\exp \left[ {i\frac{\left( {x - x'} \right)^2}{4t}} \right]$.

After some tedious, albeit straightforward calculations, the
solution for the initial wave function ($\psi _{Bi} \left( {x,t}
\right)$) can be written
\begin{widetext}
\beq \label{eq6} &&\psi \left( {x,t} \right) = \\
&&\frac{1}{2}\left\{ {e^{it}\left[ {e^{ - \left| x
\right|}\mbox{erfc}\left[ {\sqrt {it} - \frac{\left| x
\right|}{2\sqrt {it} }} \right] + e^{\left| x \right|}\mbox{erfc}
\left[ {\sqrt {it} + \frac{\left| x \right|}{2\sqrt {it} }}
\right] \frac{1 - \mu }{1 + \mu }} \right] + \frac{2\mu }{\mu +
1}e^{i\mu ^2t - \mu \left| x \right|}\left[ {\mbox{erf}\left[
{\sqrt {it} \mu - \frac{\left| x \right|}{2\sqrt {it} }} \right] +
1} \right]} \right\}\nonumber
 \eeq
\end{widetext}
 It should be stressed that Eq.\ref{eq6} is the \textit{exact} solution without any approximations.

Note that for $\mu = 1$ there is no change in the potential, and
therefore Eq. \ref{eq6} degenerates to the eigenfunction $\psi
\left( {x,t} \right) = \psi _{Bi} \left( {x,t} \right) = \exp
\left( { - \left| x \right| + it} \right)$, which means that the
wavefunction remains in its initial state.

\begin{figure}[htbp]
\centerline{\includegraphics[width=8cm,bbllx=90bp,bblly=540bp,bburx=375bp,bbury=775bp]{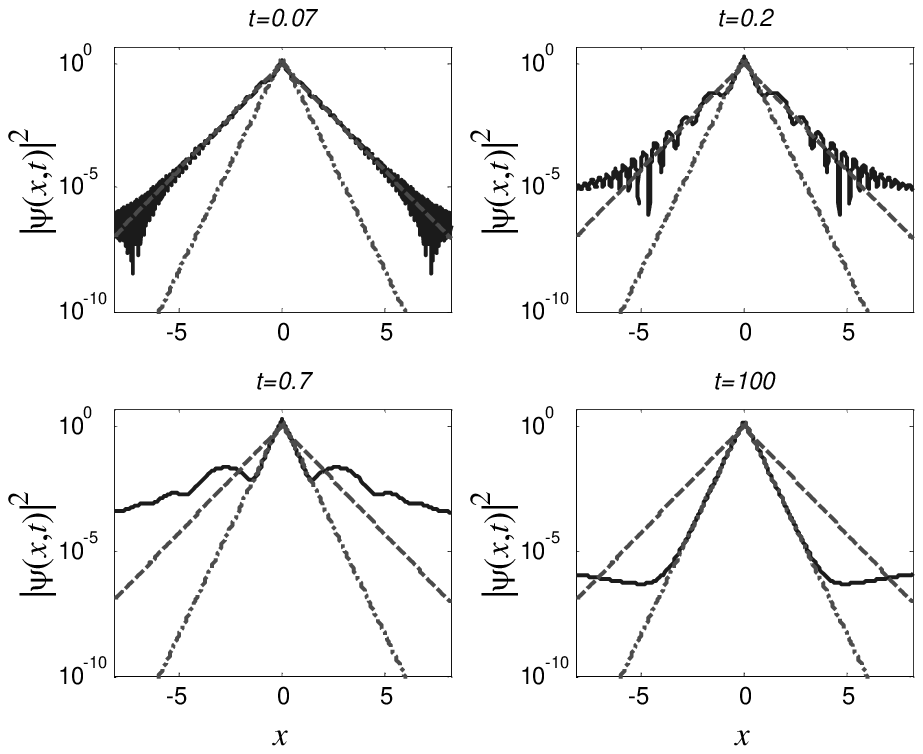}}
\caption{\emph{The distribution of the probability density in
space for four different times: $t = 0.07,0.2,0.7$ and $t = 100$.
The green dashed line represents the initial states $\left| {\psi
\left( {x,0} \right)} \right|^2$, the red dash-dotted line is the
final bound state $\left| {\psi \left( {x,\infty } \right)}
\right|^2$, and the solid blue line is $\left| {\psi \left( {x,t}
\right)} \right|^2$.}} \label{fig2}
\end{figure}

On the other hand, when $\mu = 0$, after the transition (the
abrupt change) the potential vanishes and the wavefunction
propagates freely in space
\beq \label{eq7}
 &&\psi \left( {x,t} \right) = \\
 &&\frac{1}{2} e^{it} \left\{ {e^{ - \left| x \right|}
\mbox{erfc}\left[ {\sqrt {it} - \frac{\left| x \right|}{2\sqrt
{it} }} \right] + e^{\left| x \right|}\mbox{erfc}\left[ {\sqrt
{it} + \frac{\left| x \right|}{2\sqrt {it} }} \right]} \right\}
\nonumber
\eeq

This is a relatively simple but important case (due to its generic
nature), so we would like to elaborate on its dynamics.



In this case $\left| {\psi \left( {x,0} \right)} \right|^2$ and
$\left| {\psi \left( {x,t} \right)} \right|^2$ are similar (except
for the oscillations) only till a certain $x$, beyond which the
wavefunction decays like $\left| {\psi \left( {x,t} \right)}
\right|^2\sim x^{ - 2}$ since for $x^2 / t > > 1$

$ \label{eq8} \psi \left( {x,t} \right) \cong \frac{1}{\sqrt {i\pi
t} }\frac{\exp \left( {ix^2 / 4t} \right)}{1 + \left( {x / t}
\right)^2} $.

This result is consistent with the prediction of
ref.\cite{GranotMarchewka_12} that the wavefunction at very short
times and long distances, i.e., $x^2 / t > > 1$, is

$\label{eq9} \psi \left( {x,t} \right)\sim \left[ {\psi '\left( {0
+ ,0} \right) - \psi '\left( {0 - ,0} \right)} \right]\frac{t^{3 /
2}}{\sqrt {i\pi } x^2}\exp \left( {i\frac{x^2}{4t}} \right)
$

(the initial exponentially small value of the wavefunction at $x
\to \infty $ was ignored). When $0 < \mu < 1$ the dynamics is more
intricate since the final states can be either extended (as in the
$\mu = 0$ case) or localized (Eq.2). The plot of the probability
density $\left| {\psi \left( {x,t} \right)} \right|^2$ as a
function of $x$, as depicted in Fig.1 illustrates this point.

When the perturbation is turned on the initially localized
particle's energy is modified and the particle can remain
localized at a different energy, i.e., $E_0^f = - \mu ^2$ (instead
of the initial one $E_0^i = - 1)$ but it can also escape to the
continuum.

At short times, $t < < x^2$ the wavefunction can be approximated
by
\begin{equation}
\label{eq10} \psi \left( {x,t} \right) \cong e^{it - \left| x
\right|} -  4\left( {1 - \mu }
\right)\\
\frac{\left( {it} \right)^{3 / 2}}{\sqrt \pi }\frac{\exp \left(
{i\mu ^2 / 4t} \right)}{x^2}
\end{equation}
For $x \to \infty $ the wavefunction's pertrubative term decays
like $x^{ - 2}$. In Eq.\ref{eq10} we see that the short time
behavior have a fractional power law and deviates from the
reversible $t^2$ dependence \cite{CohenTannoudjietal_8}.

\begin{figure}[htbp]
\centerline{\includegraphics[width=8cm,bbllx=90bp,bblly=540bp,bburx=375bp,bbury=775bp]{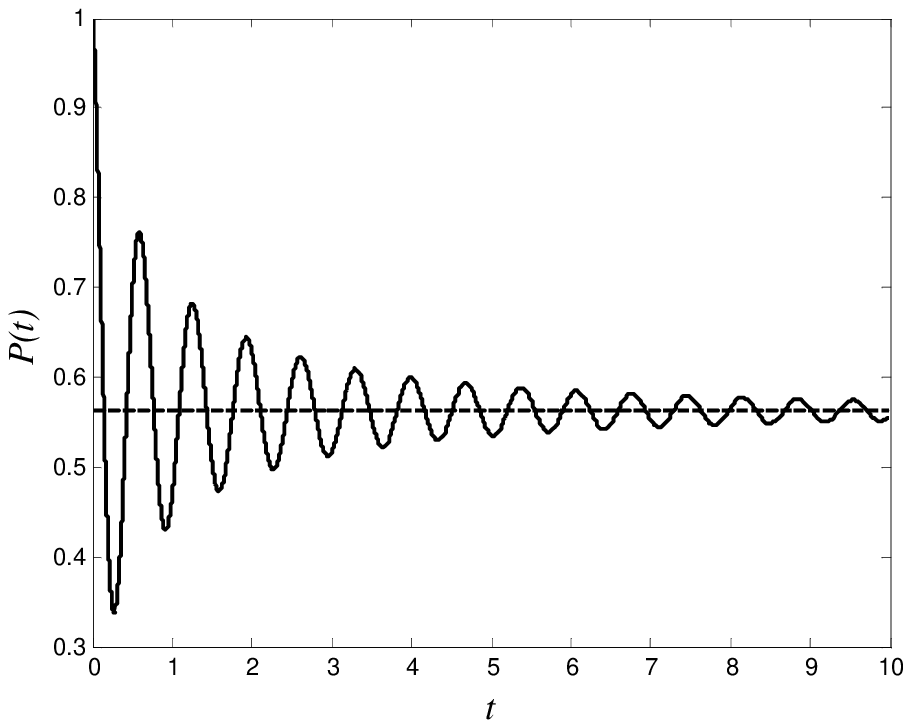}}
\caption{\emph{The temporal revolution of the probability density
for $\mu = 3$ (solid line). The dashed line stands for its final
value $P\left( {t = \infty } \right)$.}} \label{fig3}
\end{figure}

In the long time regime, i.e., $t \to \infty $, the wavefunction
can be approximated

\beqq \label{eq11}
\psi \left( {x,t} \right) \cong \left( {\mu ^{
- 1} - 1} \right)\sqrt {\frac{i}{\pi }} \left| x \right|t^{ - 3 /
2}\exp
\left( {ix^2 / 4t} \right) + \\
\frac{4\mu }{1 + \mu }\exp \left( {i\mu ^2t - \mu \left| x
\right|} \right). \eeqq

We observe two different dynamics regimes. At very large distances
from the origin (but still $t > > x)$ the first term, which is
related to the propagating-waves rules, while at short distances
the first term is merely an oscillating correction to the second
one, which is related to the final localized state. As $t \to
\infty $ the wavefunction converges to the final bound state
(\ref{eq4}) with extra factor of $4\mu ^{3 / 2} / \left( {\mu + 1}
\right)$.

Usually, the measured quantity, which quantify the decay rate, is
the survival amplitude $A\left( t \right) = \int\limits_{ - \infty
}^\infty {\psi ^\ast \left( {x,0} \right)\psi \left( {x,t}
\right)dx} $; and the survival probability $P\left( t \right) =
\left| {A\left( t \right)} \right|^2$ is the probability to remain
in the initial state (i.e., the non-decay probability); similarly,
$1 - P\left( t \right)$ is the probability to escape to infinity,
i.e., to decay.

$A\left( t \right)$ can be calculated exactly and
straightforwardly (albeit with tedious calculations). For the
initial state ($\psi _{Bi} \left( {x,t} \right)$) we find:

\begin{widetext}
\beq \label{eq11_2}
 &&A\left( t \right) = \frac{1}{\left( {1 + \mu } \right)^2}\\\nonumber
&& \left\{ {\mbox{erfc}\left( {\sqrt {it} } \right)\left[ {1 + \mu
^2 - 2it\left( {1 - \mu ^2} \right)} \right] + 2\left( {1 - \mu
^2} \right)\sqrt {\frac{it}{\pi }} e^{ - it} + 2\mu e^{it\left(
{\mu ^2 - 1} \right)}\left[ {2 - \mbox{erfc}\left( {\sqrt {it} \mu
} \right)} \right]} \right\}
 \eeq
 \end{widetext}

For completeness we add the special $\mu = 0$ case:

\begin{equation}
\label{eq12} A\left( t \right) = {\mbox{erfc}\left( {\sqrt {it} }
\right)\left[ {1 - 2it} \right] + 2\sqrt {\frac{it}{\pi }} e^{ -
it}}
\end{equation}

In Fig.2 the dynamics of the survival probability is presented for
$\mu = 3$. It is clear from the figure that the probability decays
eventually irreversibly to a constant value. However, it
\textit{never} decays exponentially.

At long time the survival amplitude goes like

\[
\label{eq13} A\left( {t \to \infty } \right)\sim \frac{4\mu
e^{it\left( {\mu ^2 - 1} \right)}}{\left( {1 + \mu }
\right)^2}\left[ {1 + \frac{\left( {\mu - \mu ^{ - 1}}
\right)^2}{4\mu \pi ^{1 / 2}\left( t \right)^{3 / 2}}e^{ - it\mu
^2 - i3\pi / 4}} \right]
\]

and the non-decay probability can be approximated

\beqq \label{eq14}&& P\left( t \right) = \left| {A\left( {t \to
\infty } \right)} \right|^2\sim \\
&&\frac{\left( {4\mu } \right)^2}{\left( {1 + \mu }
\right)^4}\left[ {1 + \frac{\left( {\mu - \mu ^{ - 1}}
\right)^2}{2\mu \pi ^{1 / 2}\left( t \right)^{3 / 2}}\cos \left(
{t\mu ^2 + 3\pi / 4} \right)} \right] \eeqq

It oscillates with angular frequency $\mu ^2$ with varying
amplitude that decays like $t^{ - 3 / 2}$ and converges to the
value $\label{eq15} P\left( t \right) \to \frac{16\mu ^2}{\left(
{\mu + 1} \right)^4}. $

It should be noted that this final probability is smaller than 1
for either $\mu > 1$ or $\mu < 1$. This result obviously
contradicts the classical intuition that the non-decay probability
decreases only when the well is raised.

Despite the irreversible nature of the process, it has no
similarity to the well-known exponential decay.

At short times $t < < 1$ this expression can be expanded by
fractional powers series to
\beqq
 &&A\left( {t < < 1} \right)\sim 1 + 2it\left( {\mu - 1}
\right) + \frac{8\left( { - 1} \right)^{3 / 4}\left( {\mu - 1}
\right)^2t^{3 / 2}}{3\sqrt \pi } - \\
&&\left( {\mu - 1} \right)^2\mu t^2 - \frac{8\left( { - 1}
\right)^{1 / 4}\left( {\mu - 1} \right)^2\left( {2\mu ^2 - 1}
\right)t^{5 / 2}}{15\sqrt \pi } + \cdots . \eeqq

\begin{figure}[htbp]
\centerline{\includegraphics[width=8cm,bbllx=90bp,bblly=540bp,bburx=375bp,bbury=775bp]{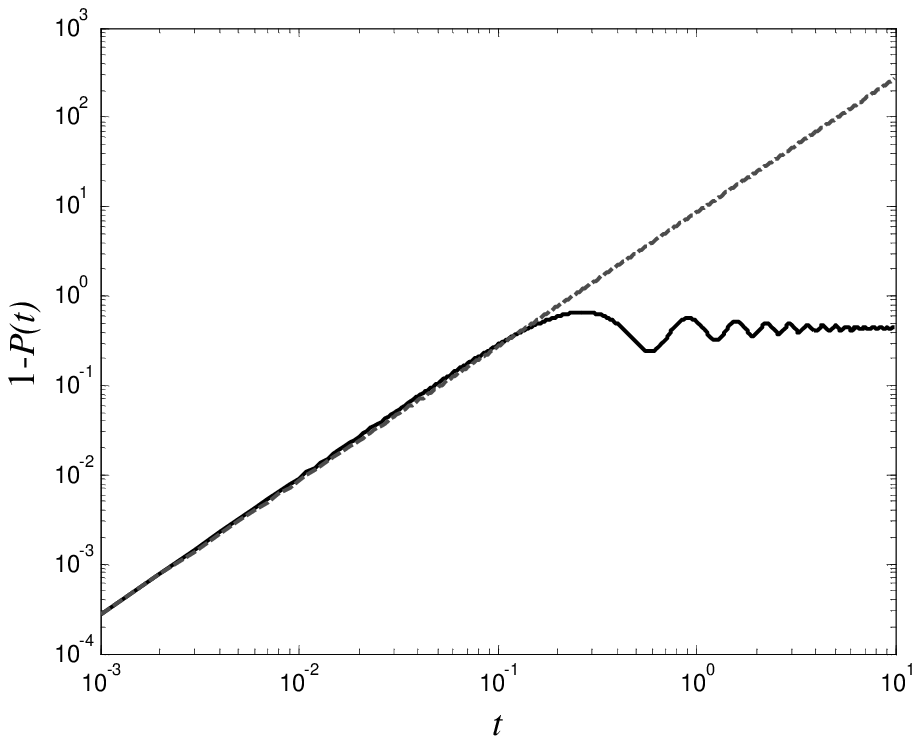}}
\caption{\emph{The escape probability $1 - P\left( t \right)$ as a
function of time (the solid black line) for $\mu = 3$. The dashed
red line stands for the short time approximation
(Eq.\ref{eq16}).}} \label{fig4}
\end{figure}

Therefore the non-decay probability goes like

\begin{equation}
\label{eq16} 1 - P\left( {t < < 1} \right) = 1 - \left| {A\left(
{t < < 1} \right)} \right|^2\sim \frac{8}{3}\sqrt {\frac{2}{\pi }}
\left( {\mu - 1} \right)^2t^{3 / 2}
\end{equation}

\noindent which means that the leading term in the escape
probability has a \textit{fractional} power dependence on time
(see Fig.3).

This behavior resembles \cite{Chiuetal_3} where the dynamics of an
ad-hoc potential spectrum was investigated; however, one of the
advantages of the model present here is its physical realization.
To the best of our knowledge, this model is the only decay model,
which allow for an exact explicit analytical solution.

The $\mu\to 1$ regime calls for comparison with perturbation
methods, which lead to the Fermi Golden Rule. In the latter case
the short time regime goes like $t^2$ instead of $t^{3 / 2}$ of
Eq.\ref{eq16}.

As was said at the beginning of the paper, it should be stressed
that even if the well had a finite width $\Delta x$ (instead of a
delta function one), then the fractional behavior, which appears
at Eqs.\ref{eq11_2}-\ref{eq16} would still be traced for
\emph{every} shallow potential well (i.e., $|E_0|\sim(U_0\Delta x
)^2/4<<\Delta x^{-2}$), and it is not merely a mathematical
anomaly.

To summarize, the dynamics of a perturbed delta function potential
well was investigated. Although this scenario can model a
realistic case (such as a particle decay from a point potential
trap), it has an \emph{exact analytical solution}. Not only does
this model behave differently than the well-known exponential
decay law as classical decay laws predict, but it does not even
have an integral power law at short times as quantum processes
suggest. In fact, the dynamics is more intricate and has a
fractional power law at short times.
\\
We believe that the analyticity of the solution of this model
along with its experimental feasibility can be used to shed light
on the generic decay process from both practical as well as
theoretical perspectives.


\begin{thebibliography}{99}

\bibitem{Fondaetal_1} L. Fonda, G.C. Ghirardi, and A. Rimini, {\em Rep. Prog.
Phys.} \textbf{41}, 587 (1978)

\bibitem{Greenland_2} P.T. Greenland, {\em Nature} \textbf{335}, 298 (1988)

\bibitem{Chiuetal_3} C.B. Chiu, E.C. G. Sudarshan, and B. Misra, {\em Phys. Rev.
D} \textbf{16}, 520-529 (1977)

\bibitem{Khalfin_4} L.A. Khalfin, {\em Zh. Eksp. Teor. Fiz.} \textbf{33} 1371 [{\em
JETP} \textbf{6}, 1053-1063 (1958)]

\bibitem{Menonetal_5} V.J. Menon and A.V. Lagu, {\em Phys. Rev. Lett.} \textbf{51}, 1407
(1983)

\bibitem{Wilkinsonetal_6} S.R. Wilkinson, C.F. Bharucha, M.C. Fischer, K. W. Madison, P.R.
Morrow, Q. Niu, B. Sundaram and M.G. Raizen, {\em Nature},
\textbf{387},  575-577 (1997)

\bibitem{Grenland_7} P.T. Greenland, {\em Nature} \textbf{387}, 548-549 (1997)

\bibitem{CohenTannoudjietal_8} See, for example, C. Cohen-Tannoudji, B. Diu and F. Laloe,
\textit{Quantum Mechanics} Vol. II (Hermann, Paris 1977)

\bibitem{Szriftgiseretal_9} P. Szriftgiser, D. Guery-Odelin, M. Arndt, and J.
Dalibard, {\em Phys. Rev. Lett.} \textbf{77}, 4 (1996)

\bibitem{Fortetal_10} C. Fort, L. Fallani, V. Guarrera, J. E. Lye, M. Modugno, D.S.
Wiersma, and M. Inguscio, {\em Phys. Rev. Lett.} \textbf{95}
170410 (2005).

\bibitem{Granot_11} E. Granot, {\em Phys. Rev. B} \textbf{71}, 035407 (2005)

\bibitem{GranotMarchewka_12} E. Granot and A. Marchewka, {\em Europhys. Lett.} \textbf{72}, 341
(2005).

\bibitem{Kleber} M.Kleber {\em Phy.Rep.} \textbf{236},331 (1994)

\bibitem{AbramowitzStegun_14} M. Abramowitz and A. Stegun, \textit{Handbook of Mathematical
Functions} , Dover publication (New-York 1965)

\end{thebibliography}
\end{document}